\documentclass[12pt,epsf,showkeys]{article}

\usepackage{graphicx,float}
\usepackage{mathrsfs,array,multirow}
\usepackage{amstext}
\usepackage{subfigure}
\usepackage{bm,latexsym,amsmath,amsfonts,amssymb}
\usepackage[usenames,dvipsnames]{color}
\usepackage{color}
\usepackage[colorlinks=true,linkcolor=blue]{hyperref}
\usepackage{soul}
\usepackage{epsfig}

\newcommand{\be}{\begin{equation}}

\newcommand{\ee}{\end{equation}}

\newcommand{\ba}{\begin{array}}

\newcommand{\ea}{\end{array}}

\begin{document}
\begin{titlepage}
\vspace{.5in}
\begin{flushright}
%CQUeST-2013-
\end{flushright}
\vspace{0.5cm}

\begin{center}
{\Large\bf Charged Rotating Black Hole with an Anisotropic Matter Field: Solution of the Maxwell Field }\\

\vspace{.4in}

  {$\mbox{Hyeong-Chan\,\, Kim}^{\P}$}\footnote{\it email: hckim@ut.ac.kr},\,\,
  {$\mbox{Bum-Hoon \,\, Lee}^{\S\dag}$}\footnote{\it email: bhl@sogang.ac.kr},\,\,
  {$\mbox{Wonwoo \,\, Lee}^{\S}$}\footnote{\it email: *warrior@sogang.ac.kr},\, \,
  {$\mbox{Youngone \,\, Lee}^{\P}$}\footnote{\it email: youngone@ut.ac.kr}\\

\vspace{.3in}

{\small \S \it Center for Quantum Spacetime, Sogang University, Seoul 04107, Korea}\\
{\small \dag \it Department of Physics, Sogang University, Seoul 04107, Korea}\\
{\small \P \it School of Liberal Arts and Sciences, Korea National University of Transportation, Chungju 27469, Korea}\\

\vspace{.5in}
\end{center}
\begin{center}
{\large\bf Abstract}
\end{center}
\begin{center}
\begin{minipage}{4.75in}

{\small \,\,\,\,
We study a rotating black hole with anisotropic matter and electromagnetic fields. 
We show that the electromagnetic field has the same form as the corresponding one in the Kerr-Newman geometry. 
The reason is thanks to the specific form of the metric function. 
We also mention energy extraction mechanisms from the black hole.}
\end{minipage}
\end{center}
\end{titlepage}

\newpage

\section{Introduction \label{sec1}}

\hspace{\parindent} A physicist would like to investigate, describe,
and understand nature, which includes the universe itself
and the astrophysical objects that are the components of the universe.
Because most astrophysical objects in the universe rotate,
it is natural for us to seek rotating solutions and investigate their properties.
However, the equation of motion under gravity is not easy to solve directly in general,
and there is no guarantee the resolution to be done straightforwardly.

Black holes are one of the most mystical and fascinating objects
in the universe~\cite{Schwarzschild:1916uq, rei1916, Nord1918}.
As we know, the first rotating black hole solution is the Kerr one~\cite{Kerr:1963ud, Kerr:2007dk}.
The second one corresponds to a Kerr-Newman black hole~\cite{Newman:1965my, Adamo:2014baa},
a charged rotating solution in asymptotically flat spacetime.
After the discoveries of these two solutions, there have been a variety of works
on finding rotating black hole solutions in various theories of gravitation \cite{Myers:1986un, Sen:1992ua, Kim:1998hc,
Gibbons:2004js, Bambi:2013ufa, Herdeiro:2014goa, Azreg-Ainou:2014pra, Toshmatov:2015npp, Kim:2019hfp, Simpson:2021dyo}.
Due to the algebraic complexity of the computations,
one has used the relevant theorem~\cite{GolSa1962}
or known algorithms\cite{Newman:1965tw, Ernst:1967wx}.

The fact that most black holes in nature coexist with dark matter or
dark energy has motivated us to study black holes that coexist
with matter fields in various theories of gravitation~\cite{Cho:2017nhx, Kiselev:2002dx, Zou:2019ays, Myung:2020dqt}.
For instance, supermassive black holes inside galaxies coexist with dark matter \cite{Becerra-Vergara:2021gmx}.
For this purpose, we obtained rotating black hole solutions that coexist with
anisotropic matter fields that generalize the Kerr-Newman one~\cite{Kim:2019hfp}.
In the solution, anisotropic matter fields are distributed continuously across
the event horizon thanks to the negative radial pressure of
those fields similarly to the electromagnetic field.

The most striking characteristic of a rotating black hole is the fact that one can extract its energy.
The rotational energy that can be extracted is located outside of the black hole event horizon.
This energy extraction mechanism is known as the Penrose process~\cite{Penrose:1969pc}.
Together with this, several energy extraction mechanisms are
known \cite{Zeldo1971, Blandford:1977ds, BDD1985, Kim:2002ei, Dadhich:2018gmh}.
Among these mechanisms, the Blandford-Znajek mechanism~\cite{Blandford:1977ds, Lee:1999se}
and the magnetized Penrose process~\cite{BDD1985, Dadhich:2018gmh}
are expected as the most promising mechanisms to describe astrophysical phenomena.

This paper is organized as follows: In Sec.\ 2, we derive rotating black hole solutions
based on the Newman-Janis (NJ) algorithm.
We write the null vectors both in the Eddington-Finkelstein coordinates and the Boyer-Lindquist coordinates.
In Sec.\ 3, we obtain the Maxwell tensor in our rotating black hole geometry.
In Sec.\ 4, we consider the energy extraction mechanism.
We summarize and discuss our results in the last section.

%===========================================================

\section{Black hole solution \label{sec2}}

\hspace{\parindent} We consider the action
\begin{equation}
I=\int d^4x \sqrt{-g}  \Big[\frac{1}{16\pi} (R-F_{\mu\nu}F^{\mu\nu})  +{\cal L}_{\rm am}\Big] + I_{b} \,,
\label{action}
\end{equation}
where $R$ is the Ricci scalar of the spacetime, $F_{\mu\nu}$ is the  electromagnetic field tensor,
$G = 1$ is for simplicity, ${\cal L}_{\rm am}$ describes effective anisotropic matter fields,
which may correspond to an extra $U(1)$ field as well as diverse dark matters,
and $I_{b}$ corresponds to the boundary term~\cite{Gibbons:1976ue, Hawking:1995ap}.
From Eq.~\eqref{action}, one obtains the Einstein equations
\begin{equation}
G_{\mu\nu}=R_{\mu\nu}-\frac{1}{2}R g_{\mu\nu}=8\pi T_{\mu\nu} \,,
\label{einsteineq}
\end{equation}
where $T_{\mu\nu}=\frac{1}{4\pi}(F_{\mu\alpha}  F_{\nu}^{\alpha}-
\frac{1}{4}g_{\mu\nu}F_{\alpha\beta}F^{\alpha\beta})-
2\frac{{\partial \cal L}_{am} }{\partial g^{\mu\nu}}+ {\cal L}_{\rm am} g_{\mu\nu}$,
and the source-free Maxwell equations
\begin{equation}
\nabla_{\nu}F^{\mu\nu} = \frac{1}{\sqrt{-g}}[\partial_{\nu}(\sqrt{-g}F^{\mu\nu})] =0 \,. \label{maxwelleq}
\end{equation}

\subsection{Static black hole \label{sec2-1}}

\hspace{\parindent} We consider a static black hole solution with the metric function as the form
\begin{equation}
ds^2 = -f(r)dt^2 + \frac{g(r)}{f(r)}dr^2 + r^2 d\Omega^2_2 \,. \label{metric}
\end{equation}

If one considers an electric field when solving the source-free Maxwell equations,
the electric charge is obtained as an integration constant.
We now choose the equation of state to consider the Maxwell field-like anisotropic matter field,
i.e. $p_r=-\varepsilon = -(\varepsilon_e + \varepsilon_{am})$ and
$p_{\theta}=p_{\phi}=\varepsilon_e+ w\varepsilon_{am}$.
We consider the energy density separately in this article, i.e. one from the
Maxwell field and the other from the additional anisotropic matter field,
unlike the paper~\cite{Kim:2019hfp}.
This matter field with $p_r=-\varepsilon$ corresponds to the matter
that the energy density is continuous across the black hole horizon, in which $g(r)=1$.

The metric function is obtained as
\begin{equation}
f(r)=1-\frac{2M}{r} +\frac{Q^2}{r^2} - \frac{K}{r^{2w}} \,, \label{metricfunction}
\end{equation}
where $M$ and $Q$ represent the ADM mass and the total charge of the black hole, respectively,
and $K$ is a constant.
The energy density is given by
\begin{equation}
\varepsilon(r) = \varepsilon_e + \varepsilon_{am} =\frac{Q^2}{8\pi r^{4}}+ \frac{r^{2w}_o}{8\pi r^{2w+2}} \,, \label{energydensity}
\end{equation}
where $r_o$ is a charge-like quantity of dimension of length and defined by $r^{2w}_o=(1-2w)K$.
There was an additional analysis for $w=1/2$ case in \cite{Cho:2017nhx}.
The negative radial pressure allows the anisotropic matter to distribute
throughout the entire space from the horizon to infinity. Therefore, the black hole can be in
static configuration with the anisotropic matter field.

\subsection{Rotating black hole \label{sec2-2}}

\hspace{\parindent} We begin with the retarded Eddington-Finkelstein coordinates
\begin{equation}
ds^2= - f(r) du^2 -2dudr + r^2 (d\theta^2 + \sin^2\theta d\psi^2)  \,. \label{st-EF}
\end{equation}
We consider the null tetrad \cite{Kinnersley:1969zza}.
The null tetrad $Z^{\mu}_{a}= \{ l^{\mu}, n^{\mu}, m^{\mu}, \bar{m}^{\mu} \}$
consists of two real null vectors, $l^{\mu}$ and $n^{\mu}$, and a pair
of complex null vectors, $m_{\mu}$ and $\bar{m}_{\mu}$.
They satisfy the relations
$l^{\mu}n_{\mu}=-1$, $m^{\mu}{\bar m}_{\mu}=1$, and
$l^{\mu}m_{\mu}=l^{\mu}{\bar m}_{\mu}=n^{\mu}m_{\mu}=n^{\mu}{\bar m}_{\mu}=0$.
The metric tensor can be expressed in terms of the null tetrad as
\begin{equation}
g^{\mu\nu}= \eta^{ab}Z^{\mu}_{a}Z^{\nu}_{b} = -l^{\mu}n^{\nu} -n^{\mu}l^{\nu} + m^{\mu}\bar{m}^{\nu} +\bar{m}^{\mu}m^{\nu} \,, \label{metricstandecom}
\end{equation}
where
\begin{displaymath}
\eta^{ab}= \left(\begin{array}{cccc}
0 & -1 & 0  &  0 \\
-1 & 0 & 0 & 0 \\
0  & 0 & 0 & 1 \\
0 & 0 & 1 & 0
\end{array}  \right) \,,
\label{metinve}
\end{displaymath}
where $a$ and $b$ enumerate null vectors.
The components of them are given by
\begin{eqnarray}
&& l^{\mu}=-\delta^{\mu}_1,~~m^{\mu}=\frac{1}{\sqrt{2} r} \left(\delta^{\mu}_2 +\frac{i}{\sin\theta}\delta^{\mu}_3\right)  \,,  \nonumber \\
&& n^{\mu}=-\delta^{\mu}_0+\frac{1}{2}f(r) \delta^{\mu}_1,~~\bar{m}^{\mu}=\frac{1}{\sqrt{2} r} \left(\delta^{\mu}_2 -\frac{i}{\sin\theta}\delta^{\mu}_3\right) \,.
\label{strEFnull}
\end{eqnarray}

By using the NJ algorithm~\cite{Newman:1965tw}, we
perform complex coordinate transformations,
\begin{equation}
u \rightarrow u'= u-ia\cos\theta \,, \quad  r \rightarrow r' = r +ia\cos\theta \,,
\label{nj01}
\end{equation}
and we get
\begin{eqnarray}
1-\frac{2M}{r} +\frac{Q^2}{r^2}- \frac{K}{r^{2w}} \Rightarrow
 1 -  \frac{2Mr-Q^2}{r^2 + a^2 \cos^2\theta} -\frac{K r^{2(1-w)}}{(r^2 +a^2\cos^2\theta)} \,,
\label{nj02}
\end{eqnarray}
where we omitted the prime.

Then the the null vectors turn out to be
\begin{eqnarray}
&& l^{\mu}=-\delta^{\mu}_1 \,, \quad
m^{\mu}=\frac{\left(ia\sin\theta(\delta^{\mu}_0-\delta^{\mu}_1)+\delta^{\mu}_2 +\frac{i}{\sin\theta}\delta^{\mu}_3\right)}{(r+i a\cos\theta)\sqrt{2}}  \,, \nonumber \\
&& n^{\mu}=-\delta^{\mu}_0+\frac{1}{2}F(r,\theta)\delta^{\mu}_1 \,, \quad
\bar{m}^{\mu}=\frac{\left(-ia\sin\theta(\delta^{\mu}_0-\delta^{\mu}_1)+\delta^{\mu}_2 -\frac{i}{\sin\theta}\delta^{\mu}_3\right)}{(r-i a\cos\theta)\sqrt{2}}  \,. \label{tetrad3}
\end{eqnarray}
One could get the rotating metric components $g^{\mu\nu}$.

The Eddington-Finkelstein form of the geometry is
\begin{eqnarray}
ds^2&=&  - F(r, \theta) du^2 - 2du dr +2a\sin^2\theta dr d\psi \nonumber \\
&-&2[1-F(r, \theta)]a\sin^2\theta du d\psi + \rho^2 d\theta^2
+ \frac{\Sigma}{\rho^2} \sin^2\theta d\psi^2 \,, \label{eddfin}
\end{eqnarray}
where $\rho^2=r^2+ a^2\cos^2\theta$, $\Sigma=(r^2+a^2)^2-a^2\triangle\sin^2\theta$, and $\triangle = \rho^2F(r, \theta)+a^2\sin^2\theta$
and the corresponding null vectors are given by
\begin{eqnarray}
&&l_{\mu} =\left[\delta^0_{\mu} - a\sin^2\theta\delta^3_{\mu} \right]  \,,~ n_{\mu}
=\left[\frac{F(r,\theta)}{2}\delta^0_{\mu} + \delta^1_{\mu} +\frac{a\sin^2\theta(2-F(r,\theta))}{2} \delta^3_{\mu}  \right] \,, \nonumber \\
&& m_{\mu} =\frac{\rho^2}{\sqrt{2} (r+ia\cos\theta)}\left[\delta^2_{\mu} + i\sin\theta \delta^3_{\mu} \right] \,, \quad
{\bar m}_{\mu} =\frac{\rho^2}{\sqrt{2} (r-ia\cos\theta)}\left[\delta^2_{\mu} - i\sin\theta \delta^3_{\mu} \right] \,.
\label{tetrad4}
\end{eqnarray}

By using coordinate transformations
\begin{equation}
du= dt - \frac{r^2+a^2}{\triangle} dr \,, \quad d\psi=d\phi - \frac{a}{\triangle} dr \,,\label{bolin}
\end{equation}
we obtain the Boyer-Lindquist form~\cite{Boyer:1966qh},
\begin{eqnarray}
ds^2&=&  - F(r, \theta) dt^2 -2[1-F(r, \theta)]a\sin^2\theta dt d\phi + \frac{\Sigma}{\rho^2} \sin^2\theta d\phi^2 + \frac{\rho^2}{\triangle} dr^2 + \rho^2 d\theta^2 \,, \nonumber \\
&=& - \frac{\triangle}{\rho^2} (dt - a\sin^2\theta d\phi)^2 + \frac{\sin^2\theta}{\rho^2}[a dt -(r^2+a^2) d\phi]^2  +  \frac{\rho^2}{\triangle} dr^2 + \rho^2 d\theta^2 \,,
\end{eqnarray}
and the corresponding null vectors are given by
\begin{eqnarray}
&&l_{\mu} =\left[-\delta^0_{\mu} + \frac{\rho^2}{\triangle}\delta^1_{\mu} + a\sin^2\theta\delta^3_{\mu} \right]  \,,~ n_{\mu}=\frac{\triangle}{2\rho^2}\left[-\delta^0_{\mu} -\frac{\rho^2}{\triangle} \delta^1_{\mu}
+ a\sin^2\theta\delta^3_{\mu}  \right] \,, \nonumber \\
&& m_{\mu} =\frac{1}{\sqrt{2} (r+ia\cos\theta)}\left[-ia \sin\theta \delta^0_{\mu} + \rho^2 \delta^2_{\mu}
+ i\sin\theta (r^2+a^2)\delta^3_{\mu} \right] \,, \nonumber \\
&& {\bar m}_{\mu} =\frac{1}{\sqrt{2} (r-ia\cos\theta)}\left[ia \sin\theta \delta^0_{\mu} + \rho^2 \delta^2_{\mu}
- i\sin\theta (r^2+a^2)\delta^3_{\mu}  \right] \,,
\label{tetrad5}
\end{eqnarray}
and
\begin{eqnarray}
&&l^{\mu} =\frac{1}{\triangle}\left[(r^2+a^2) \delta^{\mu}_0 + \triangle \delta^{\mu}_1 + a \delta^{\mu}_3  \right] \,,~ n^{\mu}=\frac{1}{2\rho^2}\left[(r^2+a^2)\delta^{\mu}_0 - \triangle \delta^{\mu}_1 + a \delta^{\mu}_3  \right] \,, \nonumber \\
&& m^{\mu} =\frac{1}{\sqrt{2} (r+ia\cos\theta)}\left[ia \sin\theta \delta^{\mu}_0 + \delta^{\mu}_2 + \frac{i}{\sin\theta} \delta^{\mu}_3  \right] \,, \nonumber \\
&& {\bar m}^{\mu} =\frac{1}{\sqrt{2} (r-ia\cos\theta)}\left[-ia \sin\theta \delta^{\mu}_0 + \delta^{\mu}_2 - \frac{i}{\sin\theta} \delta^{\mu}_3  \right] \,.
\label{tetrad6}
\end{eqnarray}

The components of the energy-momentum tensor can be obtained using $8\pi\varepsilon=e^{\mu}_{\hat{t}}e^{\nu}_{\hat{t}}G_{\mu\nu}$,
$8\pi p_{\hat{r}}=e^{\mu}_{\hat{r}}e^{\nu}_{\hat{r}}G_{\mu\nu}$,
$8\pi p_{\hat{\theta}}=e^{\mu}_{\hat{\theta}}e^{\nu}_{\hat{\theta}}G_{\mu\nu}$,
$8\pi p_{\hat{\phi}}=e^{\mu}_{\hat{\phi}}e^{\nu}_{\hat{\phi}}G_{\mu\nu}$, in which
\begin{eqnarray}
e^{\mu}_{\hat{t}}&=& \frac{(r^2+a^2,0,0,a)}{\rho\sqrt{\triangle}}\,,~~~~e^{\mu}_{\hat{r}}= \frac{\sqrt{\triangle}(0,1,0,0)}{\rho} \,, \nonumber \\
e^{\mu}_{\hat{\theta}}&=&\frac{(0,0,1,0)}{\rho}\,, \quad e^{\mu}_{\hat{\phi}} =-\frac{(a\sin^2\theta,0,0,1)}{\rho\sin\theta} \,. \label{otetrad}
\end{eqnarray}
shown in Ref.~\cite{Carter:1968ks, Azreg-Ainou:2014nra}.
We obtain physical quantities
\begin{eqnarray}
&&\varepsilon=\varepsilon_e + \varepsilon_{am} =\frac{Q^2}{8\pi \rho^{4}} +\frac{r^{2w}_o r^{2(1-w)}}{8\pi \rho^{4}}\,,~~
p_{\hat{r}}=-\varepsilon \,, \nonumber \\
&&p_{\hat{\theta}}=p_{\hat{\phi}}
	= \frac{Q^2}{8\pi \rho^4} + [\rho^{2}w-a^2\cos^2\theta] \frac{\varepsilon_{am}}{r^2} \,, \label{varepsilon}
\end{eqnarray}
in the orthonormal frame, $(e_{\hat{t}}, e_{\hat{r}}, e_{\hat{\theta}}, e_{\hat{\phi}})$. The corresponding set of the covariant tetrad is given by
\begin{eqnarray}
&&e^{\hat t}_{\mu} = \frac{\sqrt{\triangle}}{\rho} (1,0,0,-a\sin^2\theta)\,, ~~~
e^{\hat r}_{\mu} = \frac{\rho}{\sqrt{\triangle}} (0,1,0,0) \,, \nonumber \\
&& e^{\hat \theta}_{\mu} = \rho (0,0,1,0) \,, ~~~
e^{\hat \phi}_{\mu} =  \frac{\sin\theta}{\rho} (a,0,0, -(r^2+a^2))  \,.
\label{cotetrad}
\end{eqnarray}

\section{Maxwell Tensor\label{sec3}}

\hspace{\parindent} In this section, we obtain the components of the Maxwell tensor.
For the non-rotating charged black hole, the gauge field has only the component
$A_0=-Q/r$. While for the rotating charged black hole, how can we obtain the gauge field (or Maxwell field)?
One could consider this as a general relativistic version of the problem for finding
the potential of a rotating charge in the geometry of a rotating black hole.
When the electric charge rotates, the $A_{\phi}(r, \theta)$
component is induced as a non-vanishing one. Thus, $A_{\mu} = (A_0, 0, 0, A_{\phi})$.
One could take $A_0=-\frac{Qr}{\rho^2}$ according to Eq.~\eqref{nj02}.
To obtain $A_{\phi}(r, \theta)$, one could use the source-free Maxwell equations Eq.~\eqref{maxwelleq}.
However, one gets two nonlinearly coupled second-order differential equations in terms of $r$ and $\theta$.
It is too difficult to solve these equations analytically.
And also, it seems that one should first find the components of the Maxwell tensor independently,
and then one should check whether or not the tensor satisfies Maxwell equations.

We check it out in two ways as shown in Refs.\ \cite{Erbin:2014aya, Janis:1965tx, Newman:1965my}.
The first one is from Giampieri's prescription in \cite{Erbin:2014aya}
and the second one is from Refs.~\cite{Janis:1965tx, Newman:1965my}.
In this article, we describe the procedure.

Giampieri's prescription is as follows:
For the non-rotating charged one, $A_{\mu}dx^{\mu}=-\frac{Q}{r}du$ in the null coordinate.
By Giampieri's prescription for the rotating charged one, $du=du'-a\sin^2\theta d\psi$,
through the nontrivial transformation, i.e. $i d\theta =\sin\theta d\psi$,
so that $A_{\mu}dx^{\mu}=-\frac{Q r}{\rho^2}(du-a\sin^2\theta d\psi) = -\frac{Q r}{\rho^2}
(dt-\frac{\rho^2}{\triangle}dr - a\sin^2\theta d\phi )$ after removing the prime.
One could set $A_r$-term to be zero by a gauge transformation, and finally,
obtain the gauge potential in the Boyer-Lindquist coordinates as follows:
\begin{equation}
A_{\mu} = \frac{Q r}{\rho^2} \left( -1, 0, 0, a\sin^2\theta  \right) \,.
\end{equation}

We now follow the description in Refs.~\cite{Janis:1965tx, Newman:1965tw, Newman:1965my, Stephani:2003tm}.
Three complex invariants (Maxwell Newman-Penrose scalar) can be defined by the electromagnetic field tensor
as follows:
\begin{eqnarray}
\Phi_0 &\equiv& F_{\mu\nu}l^{\mu}m^{\nu} = \frac{1}{4} {\cal F}_{\mu\nu}V^{\mu\nu}\,,  \nonumber \\
\Phi_1 &\equiv& \frac{1}{2}F_{\mu\nu}(l^{\mu}n^{\nu} + \bar{m}^{\mu}m^{\nu} ) = - \frac{1}{8}{\cal F}_{\mu\nu}W^{\mu\nu}  \,,  \nonumber \\
\Phi_2 &\equiv& F_{\mu\nu} n^{\mu}\bar{m}^{\nu} = - \frac{1}{4}{\cal F}_{\mu\nu}U^{\mu\nu} \,,  \label{MNPinvariants}
\end{eqnarray}
where the six real components of $F_{\mu\nu}$ are replaced by the three complex $\Phi$'s.
The complex self-dual electromagnetic field tensor is defined as
${\cal F}_{\mu\nu} \equiv F_{\mu\nu} + i F^*_{\mu\nu}$,
the dual of the electromagnetic tensor is defined as
$F^*_{\mu\nu} \equiv \frac{1}{2} \varepsilon_{\mu\nu\alpha\beta}F^{\alpha\beta}$,
and $\nabla_{\nu}{\cal F}_{\mu\nu} =0$.
$V_{\mu\nu}=l_{\mu}m_{\nu}-m_{\mu}l_{\nu}$, $U_{\mu\nu}=-n_{\mu}{\bar m}_{\nu}+ {\bar m}_{\mu}n_{\nu}$,
$W_{\mu\nu}=-l_{\mu}n_{\nu}+n_{\mu}l_{\nu} + m_{\mu}{\bar m}_{\nu} - {\bar m}_{\mu}m_{\nu}$.

All contractions vanish except for
\begin{equation}
U_{\mu\nu} V^{\mu\nu} = 2\,,~~ W_{\mu\nu} W^{\mu\nu} =-4 \,.
\end{equation}
The complex self-dual tensor can be expanded as
\begin{equation}
\frac{1}{2} {\cal F}_{\mu\nu} = \Phi_0 U_{\mu\nu} + \Phi_1 W_{\mu\nu} +\Phi_2 V_{\mu\nu} \,.
\end{equation}
Then the Maxwell tensor could be recovered from three complex invariants as
\begin{equation}
F_{\mu\nu} = \Phi_0 U_{\mu\nu} + \Phi_1 W_{\mu\nu} +\Phi_2 V_{\mu\nu}
+ ({\bar \Phi}_0 {\bar U}_{\mu\nu} + {\bar \Phi}_1 {\bar W}_{\mu\nu} +{\bar \Phi}_2 {\bar V}_{\mu\nu} ) \,.
\label{maxfind}
\end{equation}
With this convention, the energy-momentum tensor is given by
\begin{equation}
T^{(EM)}_{\mu\nu} = \frac{1}{4\pi}(F_{\mu\alpha}F_{\nu}^{~\alpha}-\frac{1}{4}g_{\mu\nu} F_{\alpha\beta}F^{\alpha\beta})
= \frac{1}{8\pi}(F_{\mu\alpha}F_{\nu}^{~\alpha} + F^*_{\mu\alpha} F^{*\alpha}_{\nu} )
= \frac{1}{8\pi} {\cal F}^{~\alpha}_{\mu}{\bar {\cal F}}_{\nu\alpha} \,.
\end{equation}

In the Eddington-Finkelstein coordinates, they are given by~\cite{Janis:1965tx, Newman:1965my}
\begin{eqnarray}
\Phi_0 = 0 \,, \quad \Phi_1 =  \frac{Q}{2(r-ia\cos\theta)^2} \,, \quad
\Phi_2 =   \frac{1}{\sqrt{2}}  \frac{i Q a\sin\theta}{(r-ia\cos\theta)^3} \,.
\label{rorEFMNPinv2}
\end{eqnarray}
They introduced Eq.~\eqref{rorEFMNPinv2} and
used the spin-coefficient formalism~\cite{Newman:1961qr}.
They obtained $\Phi$'s by nontrivial integrations with the comment
that there is no simple algorithm for Eq.~\eqref{rorEFMNPinv2}.
In this article, we do not introduce the formalism fully.
However, we assume the formalism is still working in our black hole geometry.
In the Boyer-Lindquist coordinates, they are given by
\begin{eqnarray}
\Phi_0  =  0 \,,  \quad \Phi_1  =   \frac{Q}{2(r-ia\cos\theta)^2} \,, \quad
\Phi_2  =   0 \,.
\label{rorEFMNPinv3}
\end{eqnarray}
By plugging Eqs.\ \eqref{tetrad5} and \eqref{rorEFMNPinv3} into Eq.\ \eqref{maxfind},
the Maxwell tensor can be obtained as follows:
\begin{eqnarray}
F_{tr} &=& -F_{rt} = \frac{Q}{\rho^4} (a^2\cos^2\theta -r^2)\,,~~~
F_{t\theta} = -F_{\theta t} = \frac{Q}{\rho^4} (a^2 r \sin2\theta)\,,  \nonumber \\
F_{r\phi} & = & - F_{\phi r} =\frac{Q}{\rho^4} a\sin^2\theta (a^2\cos^2\theta - r^2) \,,~~~
F_{\theta\phi} = -F_{\phi\theta} =\frac{Q}{\rho^4} ar\sin2\theta (r^2 + a^2) \,.
\label{maxcod}
\end{eqnarray}
and
\begin{eqnarray}
F^{tr} &=& -F^{rt} = \frac{Q}{\rho^6} (r^2 - a^2\cos^2\theta)(r^2 +a^2)\,,~~~
F^{t\theta} = -F^{\theta t} = \frac{Q}{\rho^6} (-a^2 r \sin2\theta)\,,  \nonumber \\
F^{r\phi} & = & - F^{\phi r} =\frac{Q}{\rho^6} a (a^2\cos^2\theta - r^2) \,,~~~
F^{\theta\phi} = -F^{\phi\theta} =\frac{Q}{\rho^6} 2ar\cot\theta \,.
\label{maxcou}
\end{eqnarray}
The Maxwell tensor given by Eqs.\ \eqref{maxcod} and \eqref{maxcou}
satisfy the source-free Maxwell equations \eqref{maxwelleq}.
One can reversely check whether or not Eqs.\ \eqref{rorEFMNPinv2} and \eqref{rorEFMNPinv3}
are satisfied by using \eqref{tetrad3}, \eqref{tetrad6}, and \eqref{maxcod}.
The Maxwell tensor in our rotating black hole geometry
has the same form as the corresponding one in the Kerr-Newman
geometry as shown in \cite{Misner:1973prb}.
This is because the electromagnetic field does not interact directly
with the additional anisotropic matter field.
We leave the scalar quantity in this article, which is given by
\begin{eqnarray}
F_{\mu\nu}F^{\mu\nu} = - \frac{2Q^4}{\rho^8} [(r^2-a^2\cos^2\theta)^2 -4r^2a^2\cos^2\theta] \,.
\end{eqnarray}

In the asymptotic rest frame with $r \gg a$, the electric field
through $F^{{\hat a}{\hat b}}=e^{\hat a}_{\mu}e^{\hat b}_{\nu}F^{\mu\nu}$
takes the form as shown in \cite{Misner:1973prb}
\begin{eqnarray}
E^{\hat r}  = \frac{Q}{r^2} + \mathcal{O}\left(\frac{1}{r^3}\right)\,,  \quad
E^{\hat \theta} = \mathcal{O}\left(\frac{1}{r^4}\right) \,.
\end{eqnarray}
while the magnetic field takes the form
\begin{eqnarray}
B^{\hat r} = - 2\frac{Q a}{r^3}\cos\theta + \mathcal{O}\left(\frac{1}{r^4}\right)\,, \quad
B^{\hat \theta} = - \frac{Q a}{r^3}\sin\theta + \mathcal{O}\left(\frac{1}{r^4}\right) \,.
\end{eqnarray}
This is a dipole magnetic field and $\mathcal{M}=Qa$ corresponds to the magnetic moment of the black hole.

\section{Energy extraction from a rotating black hole \label{sec4}}

\hspace{\parindent} One of the most interesting properties of the rotating black hole is
the existence of the ergosphere. The ergosphere is a region bounded by the outer event horizon
and the static limit surface~\cite{Ruffini:1970sp}.
In this article, we mention two particular properties.

The first property is that particles should corotate with the rotating black hole.
When a particle reaches or crosses the boundary of the ergosphere, it can no longer stop.
Any particle that enters the ergosphere rotates around the black hole.
That can either go to the center of a black hole or exit the ergosphere.
Because that can escape in this way, the ergosphere cannot be the black hole boundary, the event horizon.
Due to this property, the shape of the capture-cross-section by light is transformed into an asymmetric one.
An interesting astrophysical phenomenon is that the shadow of a rotating black hole has
the asymmetric bright side due to this property~\cite{Bardeen:1973tla, Falcke:1999pj,
Badia:2020pnh, Lee:2021sws, Konoplya:2021slg, Junior:2021dyw, Li:2021riw, Badia:2021kpk,
Khodadi:2021gbc, Solanki:2021mkt, Li:2021ypw, Zeng:2021mok}.

The second property is that there exist negative energy states of particles
orbiting in the ergosphere for an observer at infinity,
which is related to the energy extraction mechanism, the Penrose process~\cite{Penrose:1969pc}.
The Penrose process is a process that can extract energy from a rotating black hole
by reducing the rotational energy of that.
We note that the crucial point of the Penrose process is that
the timelike Killing vector becomes spacelike in the ergoregion.
One could check the superradiant stability of the rotating black hole
with a matter field~\cite{Khodadi:2021mct}.

The observer with velocity $u^{\mu}$ measure the energy of a particle with momentum
$p_{\mu}$ as $E=-p_{\mu}u^{\mu}$.
If $u^{\mu}$ becomes spacelike, an observer at infinity measures the energy $E$ as negative,
while the local observer measures that as still positive.
Or, there exist orbits with the negative angular velocity (the counter-rotating orbits)
giving rise to the negative energy states in the ergosphere.
This is a crucial reason why there exist negative energy states of particles
orbiting in the ergosphere for an observer at infinity \cite{Tursunov:2019oiq}.
The problem is that there should be a big difference in velocities
between the incident particle and the fragmented particle
with the negative angular momentum \cite{Bardeen:1972fi}.

To extract maximum energy from a black hole, this process should take place on the event horizon.
In this case, it could be a reversible one, in which the size of the event horizon remains unchanged.
In this process, the black hole loses energy, but the size of the horizon remains the same.

In~\cite{Kim:2019hfp}, we extended the black hole mass-energy formula~\cite{Christodoulou:1970wf, Christodoulou:1971pcn}
with the anisotropic matter field as follows:
\begin{equation}
M^2 = \left[M_{I} + \frac{Q^2}{4M_{I}} - \frac{K r^{2(1-w)}_H}{4M_{I}} \right]^2 + \frac{J^2}{4M^2_{I}}  \,, \label{irremass}
\end{equation}
where $M_{I} \equiv \frac{\sqrt{r^2_H +a^2}}{2}$ denotes the irreducible mass of the black hole, which
is related to the area of the black hole horizon by $A=16\pi M^2_{I}$~\cite{Hawking:1971tu}.
The total energy of a black hole could be roughly
the sum of the irreducible mass, charge contribution, rotational energy,
and the contribution from the anisotropic matter field.
We have obtained that the efficiency of the black hole with an additional matter field is
better than that of the Kerr-Newman one.

\section{Summary and discussions \label{sec6}}

\hspace{\parindent}

We have obtained rotating black holes with an anisotropic matter field and an electromagnetic field.
In this article, we did not introduce spin-coefficient formalism fully.
However, we assumed that the formalism is still working in our black hole geometry.
The electromagnetic field in our rotating black hole geometry has the same form as that in the Kerr-Newman black hole.
The reason is thanks to the specific form of the metric function.
It would be interesting to analyze the components of the electromagnetic field when the metric functions
in front of $dt^2$ and $dr^2$ in Eq.~\eqref{metric} were not the inversely same function.

In Eq.~\eqref{metricfunction}, the term of different power originates from a different field.
one could add the term with a different power as the different solution.
In Eq.~\eqref{nj02}, one can choose the different power of $\rho^2$ in the denominator as the different solution.
In this sense, it seems that they have infinite classes of solutions.

Historically, after the discoveries of the Schwarzschild's, the Reissner-Nordstr\"om's,
and the Kerr's black holes, Newman and Janis had figured out a complex coordinate transformation
that develops a rotating black hole solution from a static one.
Newman then adopted the coordinate transformation rule with his students to the
Reissner-Nordstr\"om one and finally obtained a charged rotating black hole solution.
This algorithm is the most popular solution generating method,
which gives a rotating solution based on a static one.
In this algorithm, one gets the geometry of a rotating black hole first.
The appropriate energy-momentum tensor presenting the geometry is specified
by measuring the Einstein tensor by an observer based on proper tetrad coordinates.
In general, additional matter equations add difficulties to getting proper matter fields.
Janis and Newman had studied Maxwell equations.
In their analysis, they performed nontrivial integrals to obtain three complex invariants.

In our black hole solution, there are no independent equations for the anisotropic matter field.
Rather, we have assumed a specific form of the energy-momentum tensor
to the right-hand side of the Einstein equation.
Thus, even in the case of a rotating black hole, there are no independent matter equations of motion.
The purpose of this paper was to find out how the Maxwell field works and to take a supplement.
We leave studies on diverse applications relevant to this rotating black hole for future works.

We have also studied the properties of rotating black holes.
When the rotational effects are negligible, finding a static black hole solution
and studying their properties could be very useful in analyzing spacetime geometry.
For the case of a rotating black hole, one should be very careful
because of the algebraic complexity of the computations.
Sometimes, one could not have intuitive predictions and interpretations.
Thus, in most cases, one studies static black holes
first~\cite{Antoniou:2017acq, Lee:2018zym, Lee:2021uis, Cho:2016kpf, Cho:2020jqc}
and deals with the rotating black holes based on this static one.
Of course, the study begins with the rotating black hole
when considering some phenomena that only occurred in the rotating black hole geometry.

This article was prepared for the proceedings of the $17$th Italian-Korean Symposium
for Relativistic Astrophysics.

\section*{Acknowledgments}
H.-C. Kim and Y. Lee (NRF-2020R1A2C1009313), B.-H. Lee (NRF-2020R1F1A1075472), 
W. Lee (NRF-2022R1I1A1A01067336), and Center for Quantum Spacetime (CQUeST) of 
Sogang University (Grant No. NRF-2020R1A6A1A03047877)
were supported by Basic Science Research Program through the National Research Foundation 
of Korea funded by the Ministry of Education.
We would like to thank Wontae Kim, Hyungwon Lee, and Remo Ruffini for their helpful comments,
and Sang Pyo Kim and Jin Young Kim for their hospitality at the 17th Italian-Korean Symposium on Relativistic Astrophysics
held in both Kunsan National University for offline and CQUeST, Sogang Univrsity for online, Korea, August 2-6, 2021.

\end{document}